\documentclass[usenatbib,usegraphicx,useams]{mn2e}
\usepackage{times}
\usepackage{xspace}
\newcommand{\Ao}{Adaptive optics (AO)\renewcommand{\Ao}{AO\xspace}\renewcommand{\ao}{AO\xspace}\xspace}
\newcommand{\ao}{adaptive optics (AO)\renewcommand{\ao}{AO\xspace}\renewcommand{\Ao}{AO\xspace}\xspace}
\newcommand{\elts}{extremely large telescopes (ELTs)\renewcommand{\elts}{ELTs\xspace}\xspace}
\newcommand{\mcao}{multi-conjugate AO (MCAO)\renewcommand{\mcao}{MCAO\xspace}\xspace}
\newcommand{\xao}{extreme AO (XAO)\renewcommand{\xao}{XAO\xspace}\xspace}
\newcommand{\us}{\ensuremath{\mu}s\xspace}
\newcommand{\fpga}{field programmable gate array (FPGA)\renewcommand{\fpga}{FPGA\xspace}\xspace}
\newcommand{\fpgas}{FPGAs\xspace}
\newcommand{\hdl}{hardware description language (HDL)\renewcommand{\hdl}{HDL\xspace}\xspace}
\newcommand{\ffts}{fast Fourier transforms (FFTs)\renewcommand{\ffts}{FFTs\xspace}\xspace}
\newcommand{\shwfs}{Shack-Hartmann wavefront sensor (SHWFS)\renewcommand{\shwfs}{SHWFS\xspace}\xspace}
\newcommand{\psfs}{point spread functions (PSFs)\renewcommand{\psfs}{PSFs\xspace}\renewcommand{\psf}{PSF\xspace}\xspace}
\newcommand{\psf}{point spread function (PSF)\renewcommand{\psfs}{PSFs\xspace}\renewcommand{\psf}{PSF\xspace}\xspace}

\title[Acceleration of AO simulations using FPGAs]{Acceleration of adaptive
  optics simulations using programmable logic}
\author[A.G.~Basden et al.]{A.G.~Basden\thanks{E-mail:
    a.g.basden@durham.ac.uk}, F.~Ass\'emat, T.~Butterley, D.~Geng, C.D.~Saunter, R.W.~Wilson\\
  Centre for Advanced Instrumentation, Department of Physics, Durham
  University, South Road, Durham, DH1 3LE\\}
\begin{document}
\date{Released 2005 Xxxxx XX}

\pagerange{\pageref{firstpage}--\pageref{lastpage}} \pubyear{2005}
\label{firstpage}

\maketitle
\begin{abstract}
Numerical Simulation is an essential part of the design and
optimisation of astronomical adaptive optics systems.  Simulations of
adaptive optics are computationally expensive and the problem scales
rapidly with telescope aperture size, as the required spatial order of
the correcting system increases.  Practical realistic simulations of
AO systems for extremely large telescopes are beyond the capabilities
of all but the largest of modern parallel supercomputers.  Here we
describe a more cost effective approach through the use of hardware
acceleration using field programmable gate arrays.  By transferring
key parts of the simulation into programmable logic, large increases
in computational bandwidth can be expected.  We show that the
calculation of wavefront sensor image centroids can be accelerated by
a factor of four by transferring the algorithm into hardware.
Implementing more demanding parts of the adaptive optics simulation in
hardware will lead to much greater performance improvements, of up to
1000 times.
\end{abstract}
\section{Introduction}
\Ao is a technology widely used in optical and infra-red astronomy,
and all large science telescopes have an \ao system.  A large number
of results have been obtained using \ao systems which would otherwise
be impossible for seeing-limited observations (see for example
\citet{2004A&A...417L..21G,2005ApJ...625.1004M}).  New \ao techniques
are being studied for novel applications such as wide-field high
resolution imaging \citep{2004SPIE.5490..236M} and extra-solar planet
finding \citep{2004ASPC..321...39M}.

The simulation of an \ao system is important to determine how well an
\ao system will perform.  Such simulations are often necessary to
determine whether a given \ao system will meet its design
requirements, thus allowing scientific goals to be met.  Additionally,
new concepts can be modelled, and the simulated performance of
different \ao techniques compared (see for example
\citet{2005MNRAS.357L..26V}), allowing informed decisions to be made
when designing or upgrading an \ao system and when optimising the
system design parameters.

A full end-to-end \ao simulation will typically involve several stages
\citep{2005MNRAS.356.1263C}.  Firstly, simulated atmospheric phase screens
must be generated, to represent the atmospheric turbulence, often at
different atmospheric heights.  The aberrated complex wave amplitude
at the telescope aperture is then generated by simulating the
atmospheric phase screens moving across the pupil.  The wavefront at
the pupil is then passed to the simulated \ao control system, which
will typically include one or more wavefront sensors and deformable
mirrors and a feedback algorithm for closed loop operation.
Additionally, one or more science \psfs as seen through the \ao system
are calculated.  Information about the \ao system performance is
computed from the \psf, including quantities such as the Strehl ratio
and encircled energy.

The computational requirements for \ao simulation scale rapidly with
telescope size, and so simulation of the largest telescopes cannot be
done without special techniques, such as multiprocessor
parallelisation \citep{2004SPIE.5490..705L,2003SPIE.5169..218A} (which
can suffer from IO bandwidth bottlenecks) or the use of analytical
models \citep{2003SPIE.4840..393C} (which can struggle to represent
noise sources properly).  We here describe a different approach to
parallelisation using a massively parallel programmable hardware
architecture.  

\subsection{The Durham adaptive optics simulation platform}
At Durham University, we have been developing \ao simulation codes for
over ten years \citep{1995SPIE.2534..265D}.  The code has recently
been rewritten to take advantage of new hardware, new software
techniques, and to allow much greater scalability for advanced
simulation of \elts, \mcao and \xao systems \citep{2004SPIE.5382..684R}.

The Durham \ao simulation platform uses the high level programming
language Python to join together C, hardware and Python algorithms.
This allows us to rapidly prototype and develop new and existing \ao
algorithms, and to prepare new \ao system simulations quickly.  The
use of C and hardware algorithms ensures that processor intensive
parts of the simulation platform can be implemented efficiently.

The simulation software will run on most Unix-like operating systems,
including Linux and Mac~OS~X.  The simulation platform hardware at
Durham consists of a Cray XD1 supercomputer and a distributed cluster
of conventional Unix workstations all connected by giga-bit Ethernet.
For most simulation tasks, only the XD1 is required, though for large
models, or when multiple simulations are run simultaneously, the
entire distributed cluster can be used.

\subsection{The Cray XD1 supercomputer}
\begin{figure}
\includegraphics[width=8cm]{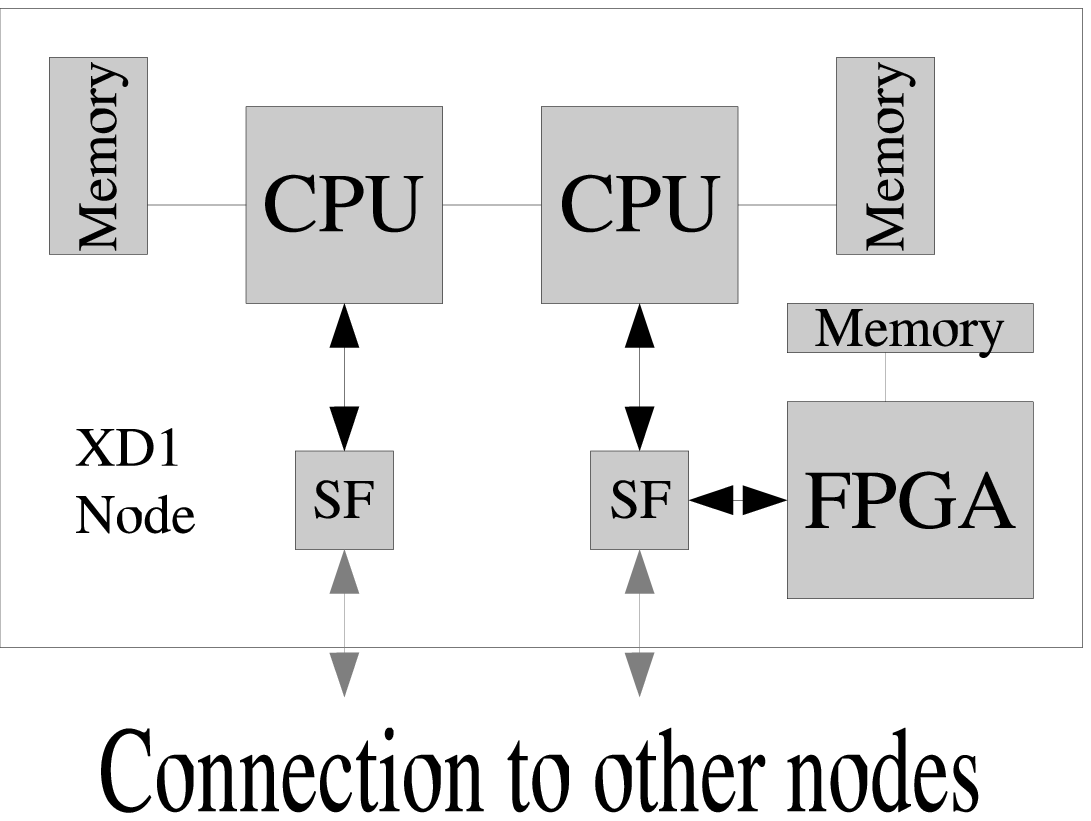}
\caption{A diagram showing an overview of the Cray XD1 hardware.  Six
  identical nodes containing two Opteron processors and an FPGA are
  connected via a high bandwidth interconnect.  In the diagram, black
  arrows represent bi-directional data transfer at 3.2~GBs$^{-1}$
  (1.6~GBs$^{-1}$ in each direction) and grey arrows represent
  bi-directional data transfer at 2~GBs$^{-1}$ (1~GBs$^{-1}$ in each
  direction).  The memory is connected with a bandwidth of
  12.8~GBs$^{-1}$ to the FPGA and 6.4~GBs$^{-1}$ to each CPU.  The
  components labelled ``SF'' are the switching fabric processors.  The
  FPGA is also connected directly to those in other nodes via
  high-bandwidth serial link.}
\label{fig:xd1}
\end{figure}

The Cray XD1 supercomputer (Fig.~\ref{fig:xd1}) is essentially a
number of dual Opteron processing nodes (running at 2.2~GHz) connected
by a high bandwidth interconnect allowing a maximum throughput of
1~GB/s between nodes in each direction \citep{crayXD1}.  At Durham, we
have a single chassis XD1, which contains six such nodes.  Within each
node, we have 8~GB of CPU memory (400~MHz DDR memory).  Each node also
contains an application acceleration module with a Xilinx Virtex-II
Pro \fpga (which allows user logic to be clocked at up to 199~MHz) and
16~MB dual port SRAM connected to the \fpga ($4\times4$~MB banks).
This \fpga is connected directly to the high bandwidth interconnect
and can be either bus master or slave.  The \fpga can access the host
Opteron memory with a bandwidth of 1.6~GBs$^{-1}$ in each direction,
without need for processor intervention when it is bus master, and
also access local SRAM memory with a bandwidth of 12.8~GBs$^{-1}$.
Additionally, a software process can write to registers or memory
within the \fpga and SRAM memory when the \fpga acts as a bus slave.

The XD1 operating system is an optimised version of Suse Linux with
improvements made by Cray \citep{crayOptimise}.  The high bandwidth
interconnect between nodes is designed for low latency communication,
with MPI having a latency of only 1.6~\us and a sustainable bandwidth
of 900~MB/s (in one direction) between nodes, as measured on the
Durham system.

\subsection{Programmable logic}
Often software algorithms will consist of a simple repetitive task
that is performed many times, thus requiring large amounts of CPU
processing time.  If this task can be offloaded from the CPU to some
application accelerator, the CPU is then free to carry out other
tasks, thus reducing the total time to complete a calculation.
Programmable logic, in the form of an \fpga is ideal for use as an
application accelerator.  A simplistic view of an \fpga is that of a
large amount of logic (AND, OR, NOT gates, latches, etc.)  which can
be connected in a user determined way.  Logic can then be built up
within the \fpga to perform simple (or even complicated) tasks.

An \fpga will normally be programmed using a \hdl, such as VHDL.  Once
written by the user, code is synthesised and mapped into the \fpga.
Common algorithms which can be placed into an \fpga include \ffts and
hardware control algorithms.

\begin{figure}
\includegraphics[width=8cm]{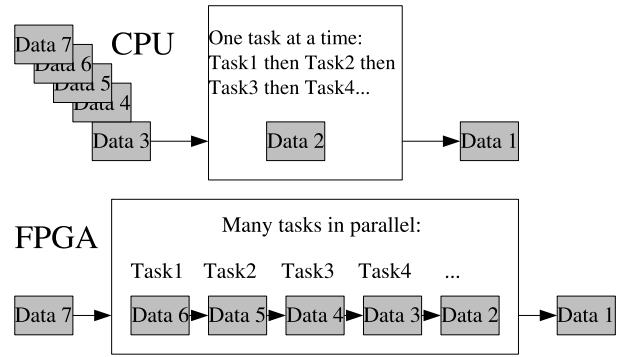}
\caption{A diagram showing a comparison of a set of calculations
  performed in serial in a CPU and the same calculations implemented
  in a parallel pipeline in an FPGA.  The CPU carries out tasks
  sequentially to a given piece of data.  The FPGA carries out many
  tasks simultaneously on different pieces of data.}
\label{fig:pipeline}
\end{figure}

An \fpga will typically be clocked at only a tenth of the speed of a
CPU.  However, by implementing many operations in parallel, they can
give a performance improvement due to the high degree of
parallelisation that they afford.  Fig.~\ref{fig:pipeline} shows a
comparison of a pipeline implemented in an \fpga and a CPU, and
demonstrates the parallelisation available to an \fpga user.

The XD1 supercomputer at Durham contains six \fpgas, each with 53,~136
logic cells.  These can be used to help accelerate the \ao
simulations.  Currently only the centroid algorithm, part of the
\shwfs has been implemented in hardware and this is able to give a
performance improvement when compared with an optimised software
implementation.  The \fpga clock speeds are user selectable, up to an
absolute maximum of 199~MHz (set by Cray).  The speed selected should
be determined dependent on the user logic within the \fpga, and \fpga
vendor tools such as ISE \citep{xilinxISE}, give an estimate for the
maximum clock speed that should be used with a given design and \fpga.

The results of the performance increases obtained from using the
\fpgas are presented in the remainder of this paper, as are details of
algorithms that will be implemented in hardware in the near future,
along with estimates of the performance improvements these will give.

\section{Hardware acceleration}
As a first step we have implemented a centroid algorithm in the
\fpgas.  Although this is not the most demanding of applications for
software, it maps well into hardware, and so is fairly straightforward
to implement.  This has allowed us to test interface code in the \fpga
which is used to read and write data from and to the host CPU memory,
develop a generic software interface to the \fpga and will also give
us some idea of the performance gains which are attainable with the
\fpga.

A \shwfs measures wavefront aberrations by measuring the wavefront
gradient across the telescope aperture.  This is done by dividing the
aperture into a grid of sub-apertures using a lenslet array.  The
centroids of each sub-aperture image then represent the local
wavefront slopes and this information can be used to shape the surface
of a deformable mirror, thus removing or reducing the wavefront
aberrations.

\subsection{Centroid algorithm}
The centroid algorithm that has been implemented in the \fpga reads
sub-aperture image photon data from the host CPU main memory
(requiring no CPU intervention), computes the $x$ and $y$ centroid
positions of this sub-aperture and writes them back into the CPU main
memory (again without CPU intervention).  The sub-aperture image
photon data is in the form of 16 bit unsigned integer numbers (typical
of CCD pixel values), and the $x$ and $y$ centroid positions are
returned in the form of 32 bit floating point numbers.  These centroid
positions are effectively the mean position of all the photons within
the sub-aperture, and are computed within the \fpga by summing each
individual photon count multiplied by its $x$ or $y$ position (in the
range 0 to $N_{x,y}-1$ where $N_{x,y}$ represents the total number of
pixels in the sub-aperture in the $x$ or $y$ direction), and dividing
by the sum of all photons to give a fixed point number which is then
converted into floating point.  Finally, an offset equal to
$\frac{N_{x,y}-1}{2}$ is subtracted from the $x$ and $y$ centroid
position so that the numbers returned to the host CPU have a range
from $-\frac{N_{x,y}-1}{2}$ to $\frac{N_{x,y}-1}{2}$ and a position of
0 represents the centre of the sub-aperture.  This \fpga
implementation gives an exact agreement with that obtained using a
traditional software centroid algorithm.

\subsubsection{User control}
User application software is used to control the hardware centroid
algorithm.  The user can specify the start address to read
sub-aperture image data from, and the number of bytes of data to read,
as well as the address to which the $x$ and $y$ centroids should be
written.  The user is also able to specify the number of pixels in the
$x$ and $y$ directions within the sub-aperture and to start and stop
the centroid calculation.  Additionally, the user can specify a pixel
weighting, which will map a 16 bit input pixel value to a weighted
pixel value to be used in the centroid algorithm.  Two weightings
commonly used are to raise the pixel values to the power of 1.5 or two.

This illustrates that although the \fpga is programmed with a fixed
function, the user code that it implements can still be programmable.
When programming an \fpga there is a trade-off between the flexibility
of the algorithm and the time taken and \fpga resources used to
implement the algorithm.

Once the centroid calculation has been started, the \fpga will read
data from the specified memory bank in host memory, and whenever
enough data has been read, will calculate a centroid value, which will be
returned to the host memory.  It is therefore possible to compute the
centroid locations of many sub-apertures with a single start command.

\subsection{Performance improvements}
The performance of the \fpga when calculating centroids has been
compared with that of optimised C code by comparing the time that each
takes to compute centroids from data in an array of given size.  This
allows us to compare performance when accessing both large and small
data arrays, and with different sized sup-apertures.

\subsubsection{Processing time}
\begin{figure}
\includegraphics[width=8cm]{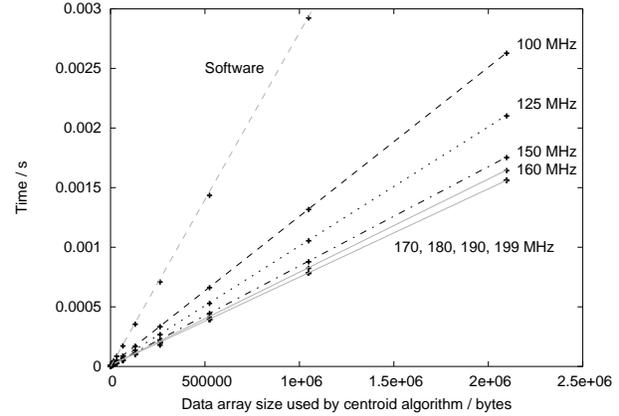}
\caption{A diagram showing the time required to apply the centroid
  algorithm to a given amount of data.  The performance of optimised C
  code (labelled software) running on the 2.2~GHz Opteron processors
  is shown along with the performance of the FPGA algorithm operating
  at different (labelled) clock speeds.  The data to which the
  centroid algorithm is applied is composed of sub-apertures each with
  16 pixels ($4\times4$), with each pixel represented by 16 bits.  The
  best fit lines are shown.}
\label{fig:time}
\end{figure}

Fig.~\ref{fig:time} shows the time taken by the \fpga and C code to
apply the centroid algorithm to a given amount of data.  Timings for
several different \fpga clock speeds are shown.  The logic we have
implemented allows us to operate the \fpga at speeds up to 199~MHz.
However, we have also investigated performance at lower clock speeds
because future developments may necessitate the use of lower clock
speeds, for example if the simulation of CCD readout noise is
implemented in the \fpga, and the logic is such that a slower clock
speed is required.

The gradient of the lines in Fig.~\ref{fig:time} represent the time in
seconds to apply the centroid algorithm to a given amount of data in
bytes.  The \fpga is able to transfer eight bytes of memory every
clock cycle both into and out of the \fpga simultaneously, and due to
the pipelined design of the \fpga centroid algorithm, data transfer
should be the performance limiting factor.  With a 100~MHz clock, the
\fpga should be able to transfer data at a maximum rate of
800~MBs$^{-1}$, corresponding to 1.25~ns per byte.  This is equal to
the gradient of the 100~MHz data on the diagram, meaning that near
maximum performance is being achieved, and that this performance is
bandwidth limited as expected.  Similarly, for clock speeds up to
170~MHz, we find that the theoretical data transfer rate is matched by
the gradients on the diagram.  However, at clock speeds above 170~MHz,
the computation time does not decrease further.

A clock speed of 199~MHz should provide a maximum data transfer rate
of 1592MBs$^{-1}$, corresponding to a data transfer time of about
0.625~ns per byte.  However, as shown in Fig.~\ref{fig:time}, this is
not achieved.

\begin{figure}
\includegraphics[width=8cm]{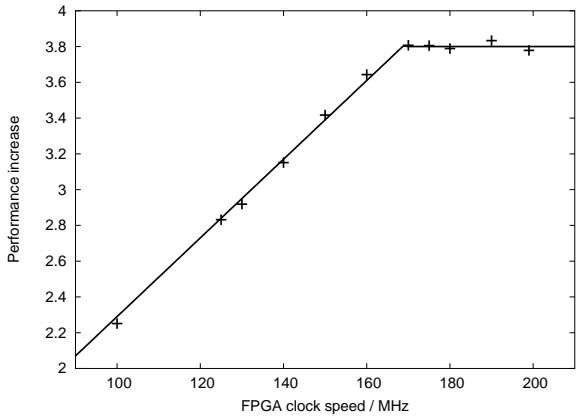}
\caption{A diagram showing the increase in performance when the
  centroid algorithm is computed in hardware instead of software as a
  function of FPGA clock speed, with the performance increase being
  the ratio of software computation time to hardware computation time.
  A sharp cutoff at 170~MHz shows the point at which the data bus is
  saturated.  A total of 2~MB of sub-aperture pixel data was
  centroided, with 16 pixels per sub-aperture and 16 bits per pixel.}
\label{fig:clkspeedup}
\end{figure}

Fig.~\ref{fig:clkspeedup} shows the performance gained when
implementing the centroid algorithm in hardware as a function of \fpga
clock speed.  The sharp cutoff at 170~MHz, corresponding to a data
transfer rate of 1.36~GBs$^{-1}$ (8 bytes per clock cycle) occurs
because this data transfer rate is approaching the theoretical maximum
allowed between the CPU and \fpga (1.592~GBs$^{-1}$).  The memory is
accessed from the \fpga via the Opteron memory controller and a fabric
switch (see Fig.~\ref{fig:xd1}), and these also have to arbitrate CPU
access to the memory and communication between other computational
nodes within the XD1 system, reducing performance below the
theoretical maximum.  This suggests that maximum data transfer
performance (and hence centroid performance) can be achieved with a
\fpga clock speed of 170~MHz.  Increasing the clock speed above this
will not give a performance increase, and the centroid data processing
pipeline in the \fpga will be idle for some clock cycles when no new
data has arrived from the memory.

\subsubsection{FPGA latencies}
\begin{figure}
\includegraphics[width=8cm]{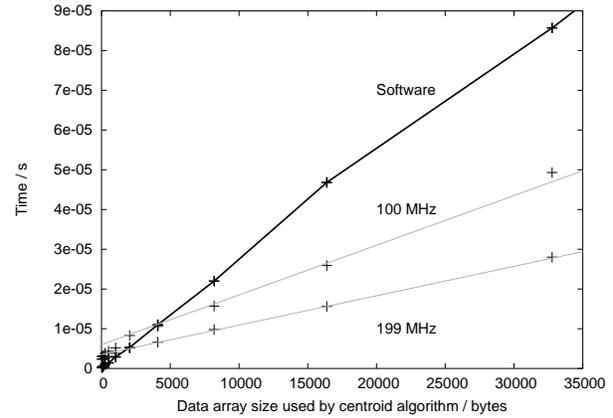}
\caption{A diagram showing the time required to apply the centroid
  algorithm to a given amount of data, as with Fig.~\ref{fig:time}.
  The solid black curve represents the time taken by the software
  implementation while the grey lines represent the time taken by the
  FPGA implementation at 100 and 199~MHz.  The grey lines are the same
  as those representing the same clock speed in Fig.~\ref{fig:time}.}
\label{fig:smalltime}
\end{figure}

Fig.~\ref{fig:smalltime} shows the time taken to apply the centroid
algorithm to a small amount of sub-aperture image photon data.  The
$y$ axis intercepts for the \fpga data are equal to about 3-5~\us
dependent on clock speed, and this represents the latencies involved
when using the \fpgas.  A functional simulation of the \fpga algorithm
(using tools provided by the \fpga vendor) shows that the time between
the last data arriving in the \fpga and leaving the \fpga should be of
order 2-3~$\mu$s (dependent on clock speed), due to the length of the
\fpga algorithm pipeline.  The additional 1-2~$\mu$s latency are due
to the memory arbitration between the CPU and \fpga and time taken by
software to start the \fpga operation.

Fig.~\ref{fig:smalltime} shows that when applying the centroid
algorithm to less than about 2-4~KB data (depending on clock speed),
it is faster to compute the centroids using the CPU, as the \fpga
overheads then begin to dominate.  However, in a typical \ao
simulation, a very large number of centroids will be computed.

\subsubsection{Relative performance}
\begin{figure}
\includegraphics[width=8cm]{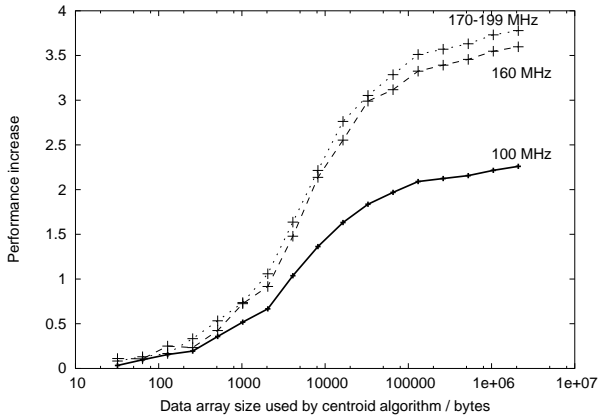}
\caption{A diagram showing the increase in performance seen when using
an FPGA centroid algorithm instead of a software algorithm.  The
performance gain is seen to level out when the centroid algorithm is
applied to more than about 100~KB of sub-aperture pixel data.  The
performance increase is defined as the ratio of the time taken by
software to the time taken by hardware when applying the centroid
algorithm.  The data to which the centroid algorithm is applied is
composed of sub-apertures each with 16 pixels, with each pixel
represented by 16 bits.}
\label{fig:speedup}
\end{figure}

The most important reason for using \fpgas in \ao simulation is for
the improved performance that they are able to give.
Fig.~\ref{fig:speedup} shows a comparison of the time taken to
calculate centroid positions using the \fpga and when using C code.

When using \fpga clock speeds above 170~MHz, the centroid algorithm is
up to four times faster in the \fpga when compared to the C algorithm for
memory blocks greater than about 100~KB.  For smaller memory blocks,
the latency introduced when using the \fpga begins to have a larger
effect and so the performance gain reduces.  However, even when using
a 4~KB memory block (two $32\times32$ pixel sub-apertures or 128
$4\times4$ pixel sub-apertures at 16 bits per pixel), the \fpga still
gives some performance gain with the software algorithm taking almost
twice as long to perform the same computation.

\begin{figure}
\includegraphics[width=8cm]{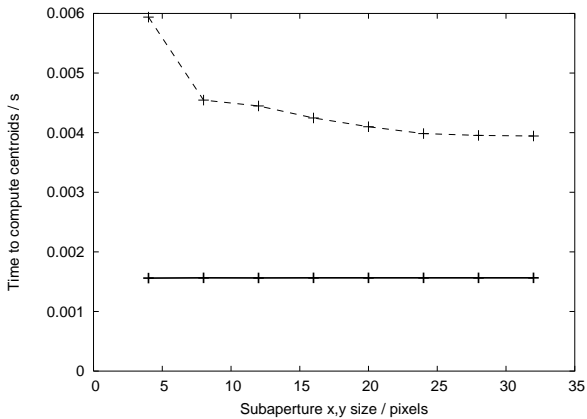}
\caption{A diagram showing the effect of sub-aperture size on centroid
  computation time for the software algorithm (dashed curve) and the
  hardware algorithm (solid line).  A total of 2~MB data were input to
  the centroid algorithm in this case (1~MPixels).}
\label{fig:subapsize}
\end{figure}

When applying the centroid algorithm to a dataset of a given size, the
performance of the software algorithm will depend partly on the number
of pixels in each sub-aperture, as shown in Fig.~\ref{fig:subapsize}.
However, the \fpga implementation does not have this dependence, since
the pipeline architecture is independent of the sub-aperture size,
with computation time depending only on the total amount of data and
the small fixed pipeline latency.

\section{Future considerations and improvements}
The performance increase of four times when using the \fpgas may seem
small, particularly when considering the amount of time required to
write and debug the VHDL code for the \fpga.  A flexible software
centroid algorithm can be written in C taking only 10--20 lines of
code.  However, when implemented in hardware, our implementation
contains about 2000 lines of code for host memory access (reading
pixel values from host CPU memory, and writing the centroid values to
host memory) and about 1200 lines of code for the centroid algorithm
(of which half of this is responsible for fixed point to floating
point conversion and pipeline flow control).  This does not include
the libraries used, which are treated as black boxes (for example Cray
specific libraries, division libraries, and \fpga block memory access
libraries), and for which, there is no source code available.

However, only about a quarter of the \fpga has been used (a large
fraction of which is the logic to read and write data from and to the
CPU memory), and there is therefore room for extra logic to perform
additional calculations.  If these additional calculations are able to
act on the data immediately before or after the centroid algorithm,
they will take virtually no extra time to compute, since the limiting
factor is the time taken to read data into and out of the \fpga.  The
performance gained over software implementations will therefore be
much greater, depending on the amount of calculation transferred into
the \fpga pipeline.

\subsection{Pipeline extensions}
\begin{figure}
\includegraphics[width=8cm]{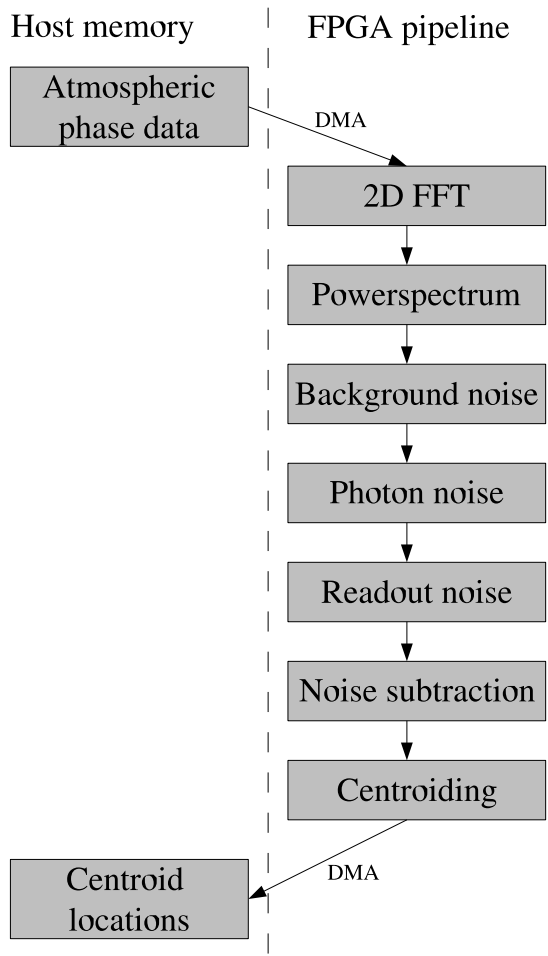}
\caption{A diagram showing the pipeline stages which will eventually
  be implemented in the FPGA logic.  Data will be read into the FPGA
  at the start of the pipeline, and the result transferred back to the
  host CPU at the end of the pipeline.}
\label{fig:future}
\end{figure}

The current success of the \fpga algorithm is thus only the start of
the hardware acceleration that can be achieved using the \fpgas and is
limited by the rate at which data can be passed into and out of the
\fpga.  We aim to extend the number of algorithms that are placed
within the hardware, including the algorithms related to the
Monte-Carlo nature of the simulation.  By ensuring that these
algorithms are part of a single \shwfs pipeline, data transfers to and
from host memory can be minimised, thus maximising the performance of
the \fpga algorithms.  The initial aim is to implement an \ao
simulation pipeline as shown in Fig.~\ref{fig:future}.  This includes
a 2D FFT of the input optical complex amplitude, high light level \psf
computation (by taking the squared modulus of the output of the 2D FFT
algorithm), the inclusion of sky background noise, photon shot noise,
CCD readout noise, noise and background subtraction and finally the
centroid algorithm.  With such a pipeline, atmospheric pupil phase
data will be read into the pipeline by the \fpga, and the centroid
values will be written back to the host CPU memory.  Due to the highly
parallel nature of \fpgas, the total computation time will be
virtually identical to that presented for the centroid algorithm here,
with a slightly higher initial latency (which is negligible when large
amounts of data are involved).  To perform such a pipeline in software
takes much longer as each stage would be performed in series, with the
total computation time being the sum of the times for each individual
stage.

\subsection{Expected pipeline performance}
Once the pipeline shown in Fig.~\ref{fig:future} has been implemented,
the time to perform these calculations in the \fpga will be slightly
greater than the times shown in Fig.~\ref{fig:time}, due to a slight
increase (of order microseconds) in the initial latency due to the
longer pipeline.  When considering the simulation of an \ao system
with $32\times32$ sub-apertures each with $8\times8$ atmospheric
complex amplitude values (32 bit floating point), a total of
$32^2\times8^2\times4=262144$ bytes will need to be read into the
\fpga, which with a 170~MHz clock will take approximately 200~\us.
The current software simulation can perform this operation in
approximately 0.25~s and so the performance will be increased by up to
1000 times, an improvement which would be difficult and expensive to
achieve using only CPU based solutions.

\subsection{Separate algorithms}
In addition to implementing the pipeline in Fig.~\ref{fig:pipeline} in
hardware, we also intend to implement several other algorithms in the
\fpga:  
\begin{enumerate}
\item Atmospheric phase screen generation
\item Wavefront reconstructor algorithms
\item Deformable mirror surface figure calculations
\item Science \psf calculation
\end{enumerate}

These algorithms will help to improve the performance of the \ao
simulations further.  Since they will not be integrated into the
\shwfs pipeline, they must operate separately from the \shwfs pipeline
and so may be placed in different \fpgas, thus having certain nodes of
the XD1 dedicated to certain parts of the \ao simulation.

\section{Conclusion}
We have presented an overview of initial performance improvements made
to the Durham \ao simulation platform using programmable logic
(\fpgas).  At present, only a centroid algorithm has been implemented
in hardware, and this gives performance improvements of up to a factor
of four when compared to optimised C algorithms.  By implementing more
of the simulation in the \fpga we expect to increase the speed of \ao
simulation by up to 1000 times.  In this way we expect to implement
realistic simulations of ELT-scale AO systems on a single Cray XD1
platform, at speeds that will be useful for detailed design and
optimisation studies of future instruments.

\bibliography{mybib} 
\bsp
\label{lastpage}
\end{document}